\documentstyle[aps,epsf,floats,preprint]{revtex}

\begin{document}

\draft

\title{Physical limits to biochemical signaling}

\author{William Bialek\footnote
{Corresponding author.  E-mail: wbialek@princeton.edu, 
Tel: (609) 258-5929, Fax: (609) 258-1549} and 
S.~Setayeshgar\footnote
{E-mail: simas@princeton.edu}
}
\address{Department of Physics, Princeton University, Princeton, New Jersey 08544}

\date{\today}

\maketitle

\begin{abstract}

Many crucial biological processes operate with
surprisingly small numbers of molecules, and there is renewed interest in
analyzing the impact of noise associated with these small numbers. 
Twenty--five years ago, Berg and Purcell
showed that bacterial chemotaxis, where a single celled 
organism must respond to small changes
in concentration of chemicals outside the cell, is limited directly by
molecule counting noise, and that aspects of the bacteria's behavioral
and computational strategies must be chosen to minimize the effects of
this noise  \cite{ref:Berg_Purcell}.  
Here we revisit  and generalize their arguments to estimate
the physical limits to signaling processes within the cell, and argue that
recent experiments are consistent with performance approaching these
limits.

\end{abstract}

\clearpage

\section{Introduction}

A striking fact about biological systems is that single molecular events can have
macroscopic consequences.  The most famous example is of course the storage of
genetic information in a single molecule of DNA, so that changes in the structure
of this single molecule (mutations) can have effects on animal behavior and body
plan from generation to generation \cite{ref:Schrodinger}.
But there also are examples where the dynamics of individual molecular
interactions can influence behavior on much shorter time scales.  Thus we (and
other animals) can see when a single molecule of rhodopsin in the rod cells of the
retina absorbs a photon \cite{ref:Rieke_98}, and some
animals can smell a single molecule of airborne odorant 
\cite{ref:olfaction}.  Even if a single molecular event does not generate a specific
behavior, it may still be that the reliability of behavior is limited by inevitable
fluctuations associated with counting random molecular events.  Thus the visual
system has a regime where perception is limited 
by photon shot noise \cite{ref:Barlow,ref:Spikes}, 
and the reliability with which bacteria can swim up a chemical gradient
appears to be limited by noise in the measurement of the gradient itself
\cite{ref:Berg_Purcell}.  It is an open question whether biochemical signaling systems
within cells operate close to the corresponding counting noise limits.

The classical analysis of bacterial chemotaxis 
by Berg and Purcell provided a simple estimate and a 
clear intuitive picture of the noise in
`measuring' chemical concentrations.  Their argument was that if we have
a sensor with linear dimensions $a$, we expect to count an average of
$\bar N \sim \bar c a^3$ molecules when the mean concentration is $\bar
c$.  Each such measurement, however, is associated with a noise $\delta
N_1\sim
\sqrt{\bar N}$.  A volume with linear dimension $a$ can be cleared by
diffusion in a time $\tau_D \sim a^2 /D$, so if we are willing to
integrate over a time $\tau$ we should be able to make $N_{\rm meas} \sim
\tau/\tau_D$ independent measurements,  reducing the noise in
our estimate of $N$ by a factor of $\sqrt{N_{\rm meas}}$.  The  result
is that our fractional accuracy in measuring $N$, and hence in measuring
the concentration $c$ itself, is given by
\begin{equation}
{{\delta c}\over {\bar c}}
= {{\delta N}\over {\bar N}} = {1\over \sqrt{\bar N N_{\rm meas}}} 
= {1\over\sqrt{D a \bar c \tau }} .
\label{BP1}
\end{equation}
A crucial claim of Berg and Purcell is that this result applies when the
sensor is a single receptor molecule, so that $a$ is of molecular
dimensions, as well as when the sensor is the whole cell,  so that 
$a\sim  1 \,\mu$m.  In particular, imagine a cell of radius $R$
which has receptor molecules of size $a$ on its surface.
With just one receptor  the limiting concentration
resolution must be $\delta c/c \sim (Da\bar c\tau)^{-1/2}$, and if $N_r$ 
receptors are distributed sparsely over the cell surface we expect
that they provide independent measurements, improving the resolution to
 $\delta c/c \sim (DN_r a\bar c\tau)^{-1/2}$.  But as $N_r$ increases to
the point where $N_r a \sim R$, this must saturate at $\delta c/c \sim
(DR\bar c\tau)^{-1/2}$, presumably because of correlations among the
concentration signals sensed by the different receptors.

The discussion by Berg and Purcell makes use of several special
assumptions which we suspect are not required, and this leads to some
clear questions:
\begin{itemize}
\item For interactions of a substrate with a single receptor, does Eq. (\ref{BP1})
provide a general limit to sensitivity, independent of molecular details?
\item Can we understand explicitly how correlations among nearby
receptors result in a limit like Eq. (\ref{BP1}), but with $a$ reflecting
the size of the receptor cluster?
\item Do the spatial correlations among nearby receptors have an analog
in the time domain, so that there is a minimum averaging time required for
noise reduction to be effective?
\end{itemize}
Finally, if we can establish Eq. (\ref{BP1}) or its generalizations as a
real limit on sensitivity for any signaling process, we would like to know
if cells actually operate near this limit. 

In most cases that we know about, biochemical signaling molecules are
thought to interact with their receptors through some kinetic process
which leads to the establishment of equilibrium between bound and
unbound states.  If this is the case, we can view the fluctuations in
occupancy of a binding site as an example of thermal noise, and we
can use the fluctuation--dissipation theorem rather than tracing through
the consequences of different microscopic hypotheses about the nature of
the interaction between signaling molecules and their targets.  We begin
with a simple example, to show that we can recover conventional results.

\section{Binding to a single receptor}

Consider a binding site for
signaling molecules, and let the fractional occupancy of the site be
$n$. If we do not worry about the discreteness of this one site, or about
the fluctuations in concentration
$c$  of the  signaling molecule, we can write a kinetic equation
\begin{equation}
{{dn(t)}\over{dt}} = k_+ c [1 - n(t)] - k_-n(t) .
\label{kinetics1}
\end{equation}
This describes the kinetics whereby the system comes to equilibrium, and
the free energy $F$ associated with binding is determined by detailed balance,
\begin{equation}
{{k_+ c}\over {k_-}} = \exp\left({F\over{k_B T}}\right) .
\label{balance}
\end{equation}
If we imagine that thermal fluctuations can lead to small changes in the
rate constants, we can linearize Eq. (\ref{kinetics1}) to obtain
\begin{equation}
{{d\delta n}\over{dt}} = -(k_+ c + k_-)\delta n + c(1 - {\bar n}) \delta
k_+ - {\bar n} \delta k_- .
\label{linear1}
\end{equation}
But from Eq. (\ref{balance}) we have
\begin{equation}
{{\delta k_+  }\over{k_+}} - 
{{\delta k_-}\over{k_-}}  = {{\delta F}\over{k_B T}}.
\end{equation}
Applying this constraint to Eq. (\ref{linear1}) we find that the
individual rate constant fluctuations cancel and all that remains is the
fluctuation in the thermodynamic binding energy $\delta F$:
\begin{equation}
{{d\delta n}\over{dt}} = -(k_+ c + k_-)\delta n + k_+ c (1-{\bar n})
{{\delta F}\over{k_B T}}.
\label{linear2}
\end{equation}
Fourier transforming,
\begin{equation}
\delta n(t) = \int {{d\omega}\over{2\pi} }\exp(-i\omega t) \,\delta{\tilde
n} (\omega) ,
\end{equation}
we can solve Eq. (\ref{linear2}) to find the frequency dependent
susceptibility of the coordinate $n$ to its conjugate force $F$,
\begin{equation}
{{\delta\tilde n (\omega)}\over{\delta\tilde F (\omega )}} = 
{1\over{k_B T}} \, 
{{k_+ c (1-{\bar n})}\over{-i\omega + (k_+ c + k_-)}}
\end{equation}
Now we can compute the power spectrum of fluctuations in the occupancy
$n$ using the fluctuation--dissipation theorem:
\begin{eqnarray}
\langle \delta n(t) \delta n (t') \rangle &=& \int {{d\omega}\over{2\pi}
}\exp[-i\omega (t-t^\prime) ] S_n (\omega)\\
S_n (\omega ) &=& {{2 k_B T }\over{\omega}}
\Im\left[{{\delta\tilde n (\omega)}\over{\delta\tilde F (\omega )}}
\right]\\
&=&
{{2 k_+ c (1-{\bar n})}\over{\omega^2 + (k_+ c + k_-)^2}} .
\end{eqnarray}
It is convenient to rewrite this as 
\begin{equation}
S_n (\omega )  = \langle (\delta n)^2\rangle
{{2\tau_c}\over{1+(\omega\tau_c)^2}} ,
\label{lorentzian1}
\end{equation}
where the total variance is
\begin{eqnarray}
 \langle (\delta n)^2\rangle &= & \int {{d\omega}\over{2\pi} }
S_n (\omega ) = k_B T {{\delta\tilde n (\omega)}\over{\delta\tilde F
(\omega )}} {\Bigg |}_{\omega = 0}\\
&=& {{k_+ c (1-{\bar n})}\over{ k_+ c + k_-}}\\
&=& {\bar n} (1-{\bar n}),
\label{variance1}
\end{eqnarray}
and the correlation time is given by
\begin{equation}
\tau_c = {1\over{k_+ c + k_-}} .
\label{tauc1}
\end{equation}
This is the usual result for switching in a
Markovian way between two states; here it follows from the
`macroscopic' kinetic equations, plus the fact that binding is an
equilibrium process.

The same methods can be used in the more general case where the
concentration is allowed to fluctuate.  Now we write
\begin{equation}
{{dn(t)}\over{dt}} = k_+ c({\vec x}_0,t) [1 - n(t)] - k_-n(t) ,
\label{kinetics2}
\end{equation}
where the receptor is located at ${\vec x}_0$, and
\begin{equation}
{\partial c(\vec x, t)\over{\partial t}}
= D \nabla^2 c(\vec x, t) - \delta (\vec x - \vec x_0 )
{{dn(t)}\over{dt}} .
\label{diffusion}
\end{equation}
Following the same steps as above, we find the linear response function
\begin{eqnarray}
{{\delta\tilde n (\omega)}\over{\delta\tilde F (\omega )}}
&=& {{k_+ c (1-{\bar n})}\over{k_B T}} \, 
{1\over{-i\omega [1 +\Sigma(\omega)] + (k_+ \bar c + k_-)}}\\
\Sigma(\omega) &=&
k_+ (1-\bar n ) \int {{d^3 k}\over{(2\pi)^3}} {1\over{-i\omega + Dk^2}}
\end{eqnarray}
The ``self--energy'' $\Sigma(\omega)$  is
ultraviolet divergent, which can be traced to the delta function in Eq.
(\ref{diffusion}); we have assumed that the receptor is infinitely
small.  A more realistic treatment would give the receptor a finite size,
which is equivalent to cutting off the $k$ integrals at some (large)
$\Lambda \sim \pi/a$, with $a$ the linear dimension of the receptor. 
If we imagine mechanisms which read out the receptor occupancy and
average over a time $\tau$ long compared to the correlation time
$\tau_c$ of the noise, then the relevant quantity is the low frequency
limit of the noise spectrum.  Hence,
\begin{equation}
\Sigma (\omega \ll D/a^2) \approx \Sigma(0) = {{k_+ (1-\bar n)}\over{2\pi D a}} ,
\end{equation}
and
\begin{equation}
{{\delta\tilde n (\omega)}\over{\delta\tilde F (\omega )}}
= {{k_+ {\bar c} (1-{\bar n})}\over{k_B T}} \,
\left[-i\omega \left(1 +{{k_+ (1-\bar n)}\over{2\pi D a}} \right) + (k_+
\bar c + k_-)\right]^{-1} ,
\end{equation}
where $\bar c$ is the mean concentration.
Applying the fluctuation--dissipation theorem once again we find the
spectral density of occupancy fluctuations,
\begin{equation}
S_n (\omega ) \approx 2 k_+ {\bar c} (1 - \bar n)
{{1 + \Sigma(0)}\over{\omega^2 (1+\Sigma(0))^2 + (k_+ \bar c + k_-)^2}} \,.
\end{equation}
We note that the total variance in occupancy is unchanged since this is
an equilibrium property of the system while coupling to concentration
fluctuations serves only to change the kinetics.  

Coupling to concentration fluctuations does serve to renormalize
the correlation time of the noise,
\begin{equation}
\tau_c \rightarrow \tau_c [ 1 +\Sigma (0)].
\end{equation}
The new $\tau_c$ can be written as
\begin{equation}
\tau_c = {{1-\bar n}\over {k_-}} + {{\bar n (1-\bar n)}\over{2\pi D a
\bar c}} ,
\end{equation}
so there is a lower bound on $\tau_c$, independent of the kinetic
parameters $k_\pm$,
\begin{equation}
\tau_c >  {{\bar n (1-\bar n)}\over{2\pi D a
\bar c}} .
\label{tauc}
\end{equation}

Again, the relevant quantity is the low frequency limit of
the noise spectrum,
\begin{eqnarray}
S_n (\omega = 0) &=&  2 k_+ {\bar c} (1 - \bar n) \cdot
{{1 + \Sigma(0)}\over{ (k_+ \bar c + k_-)^2}}\\
&=& {{2 \bar n (1 - \bar n )}\over{k_+ \bar c + k_-}}
+ {{[\bar n (1-\bar n) ]^2}\over{\pi D a \bar c}} .
\end{eqnarray}
If we average for a time $\tau$, then the root-mean-square error in our
estimate of $n$ will be 
\begin{equation}
\delta n_{\rm rms} = \sqrt{S_n (0)\cdot {1\over \tau}},
\end{equation}
and we see that this noise level has a minimum value independent of the
kinetic parameters $k_\pm$,
\begin{equation}
\delta n_{\rm rms} > {{\bar n (1-\bar n )}\over{\sqrt{\pi D a {\bar c}
\tau}}} .
\label{deltan}
\end{equation}

To relate these results back to the discussion by Berg and Purcell, we
note that an overall change in concentration is equivalent to a change in
$F$ by an amount equal to the change in chemical potential, so that
$\Delta c / \bar c \equiv \Delta F/{k_B T}$.  This means that there is an
effective spectral density of noise in measuring $c$ given by
\begin{equation}
S_c^{\rm eff} (\omega ) = \left({{\bar c}\over  {k_B T}}\right)^2 
S_F (\omega),
\end{equation}
where the `noise force' spectrum $S_F (\omega)$ is given by the
fluctuation--dissipation theorem as 
\begin{equation}
S_F(\omega ) = \Bigg| 
{{\delta\tilde n (\omega)}\over{\delta\tilde F (\omega )}} \Bigg|^{-2}
S_n (\omega ) = - {{2k_B T}\over\omega} \Im
\left[ {{\delta\tilde F(\omega )}\over{\delta\tilde n (\omega )}} \right].
\end{equation}
In the present case we find that
\begin{equation}
\label{S_ceff}
S_c^{\rm eff} (\omega ) = {{2 {\bar c} ^2}\over{k_+ {\bar c} ( 1 - \bar
n)}}\left[
1 + {{k_+ (1-\bar n)}\over{2\pi D a}}\right] .
\end{equation}
As before, the accuracy of a measurement which integrates for a time
$\tau$ is set by 
\begin{equation}
\delta c_{\rm rms} = 
\sqrt{S_c^{\rm eff} (0)\cdot {1\over \tau}},
\end{equation}
and we find  again a lower bound which is determined only by
the physics of diffusion,
\begin{equation}
{{\delta c_{\rm rms}}\over{\bar c}} > {1\over\sqrt{\pi D a\bar c \tau}}.
\end{equation}
Note that this is (up to  a factor of $\sqrt\pi$) exactly the
Berg--Purcell result in Eq. (\ref{BP1}).

\section{Binding to multiple receptors}

To complete the derivation of Berg and Purcell's original results, we
consider a collection of $m$ receptor sites at positions ${\vec x}_\mu$:
\begin{eqnarray}
{{dn_\mu(t)}\over{dt}} &=& 
k_+ c({\vec x}_\mu,t) [1 - n_\mu(t)] - k_-n_\mu(t) 
\label{kinetics3}\\
{\partial c(\vec x, t)\over{\partial t}}
&=& D \nabla^2 c(\vec x, t) - \sum_{{\rm i}=1}^N\delta (\vec x - \vec
x_\mu ) {{dn_\mu(t)}\over{dt}} .
\label{diffusion2}
\end{eqnarray}
From Eq.\ \ref{diffusion2}, we can write
\begin{equation}
	\delta c(\vec{x_\nu}, \omega) =  \frac{i \omega \Lambda}{2 \pi^2 D} \, 
              \delta \tilde{n}_\nu(\omega)
           +  \frac{i \omega}{2 \pi^2}
              \sum_{\mu \neq \nu}^m \,  
	      \frac{\delta \tilde{n}_\mu(\omega)}{\left|\vec{x} - \vec{x}_\mu\right|} \,
              \int_0^\infty \frac{k \sin{\left(k\left|\vec{x} - \vec{x}_\mu\right|\right)}}
                                    {-i \omega + D k^2} \, dk \,, 
\end{equation}
where $\Lambda$ is the cut-off wave number; as before, the cut-off arises
to regulate the delta function in Eq.\ \ref{diffusion2}, and is related to
the size of the individual receptor.
In the limit $\left(\omega/D \right)^{1/2} \ll 1$, we have
\begin{equation}
	\delta c(\vec{x_\nu}, \omega) =  \frac{i \omega \Lambda}{2 \pi^2 D} \,  
              \delta \tilde{n}_\nu(\omega)
           +  \frac{i \omega}{4 \pi D}
              \sum_{\mu \neq \nu}^m \,  
	      \frac{\delta \tilde{n}_\mu(\omega)}{\left|\vec{x} - \vec{x}_i\right|} \, ,
\end{equation}
and combining with Eq.\ \ref{kinetics3} in Fourier space, we obtain
\begin{eqnarray}
\label{eq:deltaN}
-i \omega \, \delta \tilde{N}  &=& 
   -\left[ \left(k_+ \bar{c} + k_- \right) - 
    \frac{i \omega \Lambda k_+ ( 1 - \bar{n})}{2 \pi^2 D} \right] \, \delta \tilde{N}
 \nonumber \\
 && \,\,\,\,\,  +\frac{i \omega k_+ (1 - \bar{n})}{4 \pi D} \, \sum_{\nu=1}^m \,
\sum_{\mu
\neq
\nu}  
    \, \delta \tilde{n}_\mu  
    \frac{1}{\left|\vec{x_\mu} - \vec{x}_\nu \right|} 
   + m k_+  \bar{c} \, (1-\bar{n}) \, \left(\frac{\delta \tilde{F}}{k_B T} \right) \,.
\nonumber\\
&\null&
\end{eqnarray}
where we have defined $\delta \tilde{N}(\omega) = \sum_{\mu=1}^m \delta n_\mu(\omega)$,
and assumed the steady state fractional occupancies
to be independent of the receptor site,
$\bar{n}_\mu = \bar{n} = k_+ \bar{c}/\left(k_+ \bar{c} + 
k_- \right)$ .

If we consider receptor cluster geometries such 
that the innermost sum is independent of $\vec{x}_\nu$,
we can write the sum as
\begin{equation}
\sum_{\nu=1}^m \, \sum_{\mu \neq \nu}  \, \delta \tilde{n}_\mu \, 
      \frac{1}{\left|\vec{x_\mu} - \vec{x}_\nu\right|} =  \phi(m) \cdot \delta \tilde{N} \,
\end{equation}
where
\begin{equation}
	\phi(m) = \sum_{\mu=2}^m \frac{1}{\left|\vec{x_\mu} - \vec{x}_1 \right|} \,.
\end{equation}
From the fluctuation--dissipation theorem, we find the
spectrum of $\delta \tilde{F}$ and convert that to
an equivalent concentration error as in Eq.\ \ref{S_ceff}:
\begin{equation}
	\frac{\delta c_{\rm rms}}{\bar{c}} > \frac{1}{\sqrt{\pi D \bar{c} \tau}} \, 
		\left( \frac{\Lambda}{m \pi} + \frac{\phi(m)}{2 m} \right)^{1/2} \,.
\end{equation}
For example, for receptors of radius $b$ uniformly
distributed around a ring of radius $a > b$,
we have $\phi(m) = m g_0/a$, where $g_0$ is a geometric factor of order unity, and
\begin{equation}
\label{error_cluster}
\frac{\delta c_{\rm rms}}{\bar{c}} > \frac{1}{\sqrt{\pi D \bar{c} \tau}} \, 
		\left( \frac{1}{m b} + \frac{g_0}{2 a} \right)^{1/2} \,.
\end{equation}

In summary, we find that the simple formula in Eq. (\ref{BP1}) really
does provide a general limit on the precision of concentration
measurements by sensors of linear dimension $\sim a$, at least in those
cases where the interactions between the receptor and its ligand are
passive.  Further, there is a minimum level of receptor occupancy noise
from Eq. (\ref{deltan}), and a minimum correlation time from Eq.
(\ref{tauc}).  Let us look at two examples to see how these limits
compare with the performance of real cellular signaling mechanisms.

\section{Physical examples}

\subsection{Regulation of gene expression in bacteria}  

Expression of genes is
controlled in part by the occupancy of promoter sites adjacent to the
regions of DNA which code for protein \cite{ref:Ptashne}.  We thus can view
gene expression as a readout mechanism for sensing promoter site
occupancy, or even as a sensor for the concentration of the
transcription factor proteins which bind to the promoter site.  In a
bacterium like {\em E. coli,} transcription factors are present in
$N_{\rm TF}\sim 100$ copies in a cell of volume of
$\sim 1\,\mu{\rm m}^3$
\cite{ref:Gupta_95}.  If the transcription factor is a repressor then gene
expression levels are determined by $1-n$, while if it is an activator
then expression is related to $n$; because $\delta n_{\rm rms} \propto
{\bar n} (1-\bar n )$ [Eq. (\ref{deltan})], fractional fluctuations in
either $A = n$ or $A = 1-n$ are determined by
\begin{equation}
{{\delta A}\over{\bar A}} = (1- \bar A) {1\over\sqrt{\pi D a
\bar c \tau}}
\end{equation}
Direct measurements of diffusion constants for small proteins in the {\em
E. Coli} cytoplasm yield $D\sim 3 \,\mu{\rm m}^2/{\rm s}$
\cite{ref:Elowitz_99}. A promoter site itself has linear dimensions
$a \sim 3\,{\rm nm}$, and putting these factors together we find the
crucial combination of parameters $\pi D a {\bar c} \sim 3 \,{\rm
s}^{-1}$.  In particular this means that the fluctuations in occupancy of
the promoter site, averaged over a time $\tau$, are given by
\begin{equation}
{{\delta A}\over{\bar A}} > (0.1)\cdot  (1 -{\bar A})
\cdot \left( {{100}\over N_{\rm TF}}\right)^{1/2} \cdot \left(
{30\,{\rm s}}\over\tau\right)^{1/2}
\end{equation}
Recent experiments \cite{ref:Elowitz_02} indicate that {\em E. Coli}  
achieves $\sim 10\% $ precision in control of gene expression at small
values of $\bar A$.  For
this performance to be consistent with the physical limits, the
transcription machinery  must therefore integrate the promoter site
occupancy for times of order one minute, even assuming that the
translation from occupancy to expression level itself is noiseless.
This integration can be provided by the lifetime of the mRNA transcripts
themselves, which is $\sim 3 \, {\rm min}$ in prokaryotes \cite{ref:mRNA_02}.

\subsection{Control of the flagellar motor by CheY}  
The output of bacterial
chemotaxis is control of the flagellar rotary motor \cite{ref:Berg_txt}.  
The phosphorylated form of the signaling protein 
CheY (CheY--P) binds to the motor and
modulates the probability of clockwise versus counterclockwise rotation
\cite{ref:Falke}.  Recent measurements
\cite{ref:Cluzel_00} show that the probability $p$ of clockwise rotation
depends very steeply on the concentration $c$ of CheY--P,
\begin{equation}
p ={{ c^h}\over{c^h + c_{1/2}^h}},
\label{motorprob}
\end{equation}
with $h \sim 10$ and $c_{1/2} \sim 3\,\mu{\rm M}$.  Motors switch between
clockwise and counterclockwise rotation as a simple random telegraph process, and
for $c\approx c_{1/2}$ the switching frequency is $f \approx 1.5\,{\rm
s}^{-1}$.  If we view the motor as a sensor for the internal
messenger CheY, then the observed behavior of the motor determines an
equivalent noise level of
\begin{equation}
\delta c_{rms} = \left( {{\partial p}\over{\partial c}} \right)^{-1}
\sqrt{p(1-p)} \cdot \left(\tau_0 \over \tau \right)^{1/2} ,
\end{equation}
where $\tau_0$ is the correlation time of the motor state; for the simple
telegraph model it can be shown that $\tau_0 = 2 p(1-p)/f$.  Using Eq.
(\ref{motorprob}) we find
\begin{equation}
{{\delta c_{rms}}\over c} = {1\over h} \sqrt{2\over{f\tau}} .
\end{equation}
Thus, for $c\approx c_{1/2}$, a single motor provides a readout of CheY--P
concentration accurate to $\sim 10\%$ within two seconds. 
Given the dimensions of the flagellar motor's C ring, $a \sim 45 \, {\rm nm}$, with
$m \sim 34$ individual subunits to which
the CheY-P molecules bind \cite{ref:Thomas}, 
from Eq.\ \ref{error_cluster} we find 
\begin{equation}
	{{\delta c_{\rm rms}}\over c} \sim \frac{1}{22} 
		\left( \frac{2 \, {\rm s}}{\tau} \right)^{1/2} \,, 	
\end{equation}
where we have taken the size of the individual receptor binding
site to be $b \sim 1 \, {\rm nm}$, and $D \sim 3 \, \mu {\rm m}^2/{\rm s}$
as above.  Hence, the collection of receptors
comprising the motor are able to measure the CheY-P concentration
with $\sim 5 \%$ precision within two seconds.  This is in agreement with our
earlier result obtained by observing the switching statistics of the motor
to within a factor of two.

\section{Concluding remarks}

In conclusion, we have derived from statistical mechanics considerations
the physical limits to which biological sensors that rely on the
binding of a diffusing substrate can measure its concentration.  Our
approach complements and extends the classic work by Berg and
Purcell.  For a single receptor, we arrive at their earlier
intuitive result, which states that the accuracy in measurement
of concentration is limited by the noise associated with the arrival of discrete
substrate molecules at the receptors.  Our approach extends 
in a straightforward way
to multiple receptors without relying on additional considerations;
for this case, our result demonstrates more transparently
the role of multiple receptors in improving the measurement accuracy,
as well as the saturating limit in this improvement 
set by the receptor cluster size.  
Relevant internal or external signaling molecules
are often present in low copy numbers, and their concentration
in turn regulates downstream biochemical networks crucial to
the cell's functions.  For two experimentally well-studied
examples, we show that the cell's performance is consistent
with the counting noise limits in measuring
the concentration of these signaling molecules.


\end{document}